\documentclass[prl,11pt]{article}
\usepackage{graphicx}

\newcommand{\beq}{\begin{equation}}
\newcommand{\eeq}{\end{equation}}
\newcommand{\be}[1]{\begin{equation}\label{#1}}
\newcommand{\ee}{\end{equation}}
\newcommand{\continue}{\nonumber \\ }
\newcommand{\bea}{\begin{eqnarray}}
\newcommand{\eea}{\end{eqnarray}}

\newcommand{\bm}[1]{\mbox{\boldmath{$ #1$}}}
\newcommand{\Tr}{\mbox{Tr}}

\newcommand{\pde}{\partial}
\renewcommand\Im{{\rm Im\,}}

\renewcommand{\det}{\mbox{\rm det}}

\newcommand{\MatrixII}[4]{
   \pmatrix{ {#1}  &  {#2} \cr
             {#3}  &  {#4} \cr} }

\newcommand{\FIG}[4]{\begin{figure}
		      \hspace*{0.0\textwidth}%
		      \begin{minipage}[l]{0.90\textwidth}
		      \noindent{#1}
                      
		      \caption[#2]{#3}
	              
                      \label{#4}
		      \end{minipage}
		      \end{figure} }

\newcommand{\refeq}  [1] {(\ref{#1})}
\newcommand{\reffig} [1] {fig.~\ref{#1}}
\newcommand{\refappe}[1] {appendix~\ref{#1}}

\begin{document}

\topmargin 0.5cm
\headheight 0.5cm
\title{Wave chaos in the elastic disc}

\author{ 
Niels Sondergaard 
\footnote{e-mail: niels.sondergaard@nottingham.ac.uk}
and Gregor Tanner 
\footnote{e-mail: gregor.tanner@nottingham.ac.uk}\\
School of Mathematical Sciences\\
University of Nottingham\\
University Park, Nottingham NG7 2RD, UK\\
}

\maketitle

\begin{abstract}
The relation between the elastic wave equation for plane, isotropic bodies
and an underlying classical ray dynamics is investigated. 
We study in particular the eigenfrequencies of an elastic disc 
with free boundaries and their connection to periodic rays inside the
circular domain.
Even though the problem is separable, wave
mixing between the shear and pressure component of the wave field at the
boundary leads to an effective stochastic part in the ray dynamics.
This introduces phenomena typically associated with classical chaos as for 
example an exponential increase in the number of periodic orbits. 
Classically, the problem can be decomposed into an integrable part and a 
simple binary Markov process.
Similarly, the wave equation can in the high frequency limit be mapped onto
a quantum graph. 
Implications of this result for the level statistics are discussed.
Furthermore, a periodic trace formula is derived from the scattering matrix
based on the inside-outside duality between eigen-modes and scattering 
solutions and periodic orbits are identified by Fourier transforming 
the spectral density.

\noindent
{ \normalsize submitted to {\em Physical Review E};\\
Version: 6th August 2002}
\end{abstract}

\section{Introduction}

In the beginning of the 20th century, Debye studied the density of
vibrational modes in a solid body in the context of his work on
the heat capacity. He found that the average density is in leading
order proportional to the volume of the body times the third power
of the frequency. Corrections to Debye`s result were found later
involving contributions due to the surface of the body
\cite{SAFAROV,Yale}. The density of eigenfrequencies of a solid
body contains apart form these smooth terms also oscillatory
contributions, which build up the discrete spectrum of individual
eigenmodes. These oscillatory corrections have been studied
intensively over the last decade or so in the context of the
Helmholtz and the Schr\"odinger equation. In the high frequency 
limit they are known to be  related to periodic orbits of an
underlying classical dynamics, that is, the ray dynamics in a
billiard in the former or the Hamiltonian dynamics of the
corresponding classical system in the latter case
\cite{gutbook,brack}. It was in particular observed that different
formulas apply when the classical dynamics is integrable
\cite{BerTabor} or chaotic \cite{Gutzwiller71}.
The relation between the wave equation and a related deterministic
ray-dynamics is less obvious in elasticity. The wave equations
are vectorial and different wave modes with differing wave
velocities coexist. The notion of chaos or integrability 
needs to be reexamined here, which is the main purpose of this paper.\\

In what follows we shall assume that the elastic body consists of
an isotropic material. To reduce the dimensionality of the
problem, we will furthermore consider only bodies of the form of a
thin plate or an infinite rod with constant cross section. The
vibrations in the plate or rod decouple then into two classes, the
in-plane and the anti-plane vibrations \cite{lAndl};  (for plates,
this is only true as long as the wavelength is much
smaller than the thickness of the plate). The problem is thus
reduced to two spatial dimensions. We will focus here on the
in-plane vibrations for which the wave equation is still vectorial
which makes it more complex than say the scalar Helmholtz
equation. The wave field can be decomposed into two polarizations,
that is, pressure and shear waves, which have different wave
speeds. The two polarizations couple at the boundary for
physically relevant boundary conditions. An underlying ray
dynamics emerging at high frequencies has similarly two types of rays
traveling at different speeds and conversion between
polarizations take place at the boundary. Ray conversion
introduces a stochastic component into the dynamics and may lead
to a large increase of possible ray trajectories compared to
deterministic billiards for the same domain shapes \cite{biswas}.

We shall discuss mainly the case of a circular disc here, a
separable problem due to the  spherical symmetry. The case of
elastic bodies without symmetries and fully chaotic classical ray
dynamics has been discussed in \cite{couch} including a comparison 
with the semiclassical quantization of chaotic systems 
\cite{gutbook}. The scattering from two circular cavities in 
an elastic medium has been treated in \cite{cavity}.
The common idea is to write the spectral density as the trace of the
Green's function which can in turn be expressed as sum over
classical periodic orbits. We shall derive such a trace formula
for the elastic disc and compare the results with the numerically 
calculated
spectrum. We will furthermore show that the wave equation as well
as the classical ray dynamics still posses a degree of
`randomness' due to the wave mixing at the boundary even though
angular momentum is conserved.

The quantum spectra of systems whose classical dynamics is chaotic
has been found to follow random matrix theory originally developed
in nuclear physics, see for example \cite{Meh91,Guhr98}. In the elastic case 
spectral statistics coinciding with RMT has been observed in 
experiments for rectangular and stadium shaped plates 
\cite{sornette,oxborrow,schaadt}. Recently, spectra of
graphs have also be shown to behave quite similar to chaotic
systems \cite{Kot97,Tan00}; they posses a trace formula for the spectral
density and show random matrix statistics. We will make a
connection between the ray dynamics in an elastic disc and the
dynamics on a simple Markov graph, and will show how mode
conversion effects the correlations in the eigenfrequency spectrum
of the disc.

The paper is organized as follows; we shall first introduce the
elastic wave equation and a high frequency approximation
of its boundary element kernel in section \ref{sec:sec1}. Next, the
classical ray dynamics in a disc is  discussed in section
\ref{sec:sec2a}. In section \ref{sec:sec2}, the exact solutions of
the wave equation for circular symmetries is derived and high
frequency approximations are discussed. We will then study the
so-called nearest neighbor spacing distribution for the disc
spectrum in more detail. An expression for the oscillatory part 
of the level density in terms of periodic orbits will be 
derived from the scattering matrix in section \ref{sec:sec3}.

\section{The elastic wave equation and short wavelength approximations}
\label{sec:sec1}
We shall consider the propagation of elastic deformations through an
isotropic body. The partial differential equation in the frequency domain is
the linear Navier-Cauchy equation \cite{lAndl,bedford}
\beq
\mu \Delta({\bf u}) + (\lambda + \mu) {\bf \nabla ( \nabla \cdot u)} +
\rho \omega^2 {\bf u} = 0 \,,
\label{pde}
\eeq
where ${\bf u}({\bf x})$ is the displacement field in the body,
$\lambda$, $\mu$ are the material dependent Lam\'e coefficients
and $\rho$ is the density. We shall
restrict ourselves to two-dimensional problems in what follows.
A generalization of the results in this section to three dimensions is,
however, straightforward. The two-dimensional wave equation describes
in-plane deformations in plates or wave
propagation in cylindrical bodies extending to infinity along one
axis.

Introducing elastic potentials $\Phi$ and $\bf \Psi $ by using standard
Helmholtz decomposition of the displacement field $\bf u$, that is,
\be{helm}
{\bf u} = {\bf u}_p + {\bf u}_s
\quad \mbox{with} \quad
{\bf u}_p = \nabla \Phi, \;\;
{\bf u}_s =  {\bf \nabla} \times {\bf \Psi} \; ,
\ee
the Navier-Cauchy reduces to two Helmholtz equations for the potentials
\be{Helmeqn} (\Delta + k_p^2) \Phi = 0\, ; \quad
             (\Delta + k_s^2) {\bf \Psi} = 0\, .
\ee
Here, $k_p$ and $k_s$ are the wave numbers for the pressure (or longitudinal)
and shear (or transversal) wave component, respectively. The wave velocities
relating the wave numbers to the frequency $\omega$ via the
dispersion relation $k_{p,s} = \omega/c_{p,s}$ are different for the
two different polarizations, on obtains
\[
c_p=\sqrt{\frac{\lambda + 2 \mu}{\rho}} \hspace{1cm}
c_s=\sqrt{\frac{\mu}{\rho}} .
\]
Note that the pressure wave speed is always larger than the shear wave
velocity.  It is this difference in wave speed which leads to the
phenomenon of wave splitting in the ray dynamics on impact with a boundary,
see \reffig{platefig}.
The two wave equations \refeq{Helmeqn} couple at the boundary, the details
of the coupling depend on the boundary conditions. We shall in the following
always consider free boundaries, that is, no forces act on
the surface of the body. Forces acting on general surface elements
are described in terms of the stress tensor
\[
\sigma_{ij} = \lambda \, \partial_k u_k \delta_{ij} +
\mu  \left(\partial_i u_j + \partial_j u_i\right)
\]
where the summation convention is used. Free boundary conditions correspond to
\be{freebound}
{\bf t}({\bf u}) = {\sigma}({\bf u}) \cdot {\bf n} = {\bf 0}
\ee
for the displacement field at the boundary where $\bf n$ denotes the normal to
the boundary. The operator $\bf t$ refers to the traction.
The traction operator (or traction matrix after representing it in a
particular basis) for a circular boundary will be needed later to
calculate the eigenfrequency spectrum of an elastic disc,
and is explicitly derived in \refappe{s-resoExact}. \\

\FIG{
\rotatebox{-90}{\includegraphics[height=6cm]{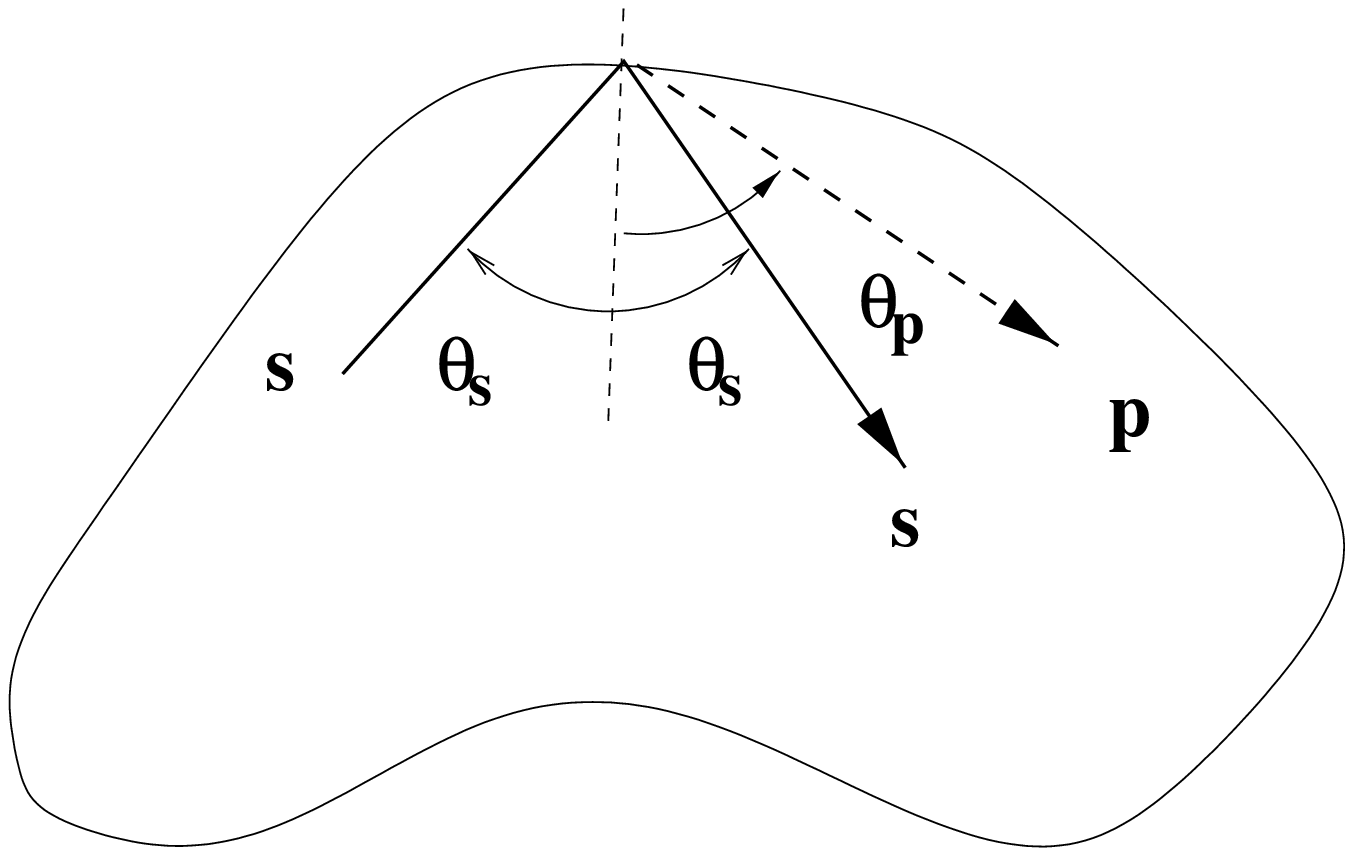}}
}{platefig}{Wave splitting for in-plane waves at the boundary}{platefig}

Waves propagate freely inside the medium, that is, the pressure and shear
component are decoupled and travel along straight lines. Wave splitting occurs
at the boundary according to Snell's law
\be{Snell}
\frac{c_p}{c_s}=\frac{\sin \theta_p}{\sin \theta_s},
\ee
where $\theta_p$, $\theta_p$ denote the angle of incident or reflection
of the pressure and shear wave, respectively, measured with respect to the
normal to the surface, see \reffig{platefig}. No mode conversion takes
place for $s$-waves coming in at incident angles larger than a critical
angle $\theta_c = \arcsin(c_s/c_p)$.  The reflection coefficients
at impact with a plane interface for free boundary conditions can be given
in terms of an orthogonal $2\times 2$ coefficient matrix $\bm{\alpha}$,
\cite{couch}
\begin{eqnarray} \label{refl-coef}
\alpha_{pp}
&=&\frac{\sin 2 \theta_s \, \sin 2 \theta_p - \kappa^2 \cos^2 2\theta_s}
            {\sin 2 \theta_s \, \sin 2 \theta_p + \kappa^2 \cos^2 2\theta_s}\\
\nonumber \alpha_{ss} &=& \alpha_{pp}\\
\nonumber \alpha_{ps} &=& - \alpha_{sp} \quad \mbox{and} \quad
\nonumber \alpha_{pp}^2 + \alpha_{ps}^2 = 1
\end{eqnarray}
where $\alpha_{\pi\pi'}$ relates an incoming wave of polarization $\pi\in \{s,p\}$
to an outgoing wave of polarization $\pi'$ and $\kappa = c_p/c_s$.
In the literature, often only the reflection coefficients 
for the displacement field ${\bf u}$ are given 
\cite{lAndl} related to the
coefficient matrix $\alpha$ above by
\beq
a_{\pi\pi'} = \sqrt{\frac{c_{\pi} \cos \theta_{\pi}}
{c_{\pi'} \cos \theta_{\pi'}}} \, \alpha_{\pi \pi'} \, .
\label{refl-coefFromS}
\eeq
Here, $|a_{\pi \pi'}|^2$ is equivalent to the proportion of the energy density of the wave function
undergoing transition from $\pi' \to \pi$ whereas $|\alpha_{\pi \pi'}|^2$ is the ratio of the 
corresponding energy fluxes normal to the boundary (with normal velocity $c_\pi \cos \theta_\pi$).
The unitarity of $\bf \alpha$ thus implies flux conservation normal to the boundary. The 
tangential energy flux is, however, not conserved for free boundary conditions due to the 
non-vanishing tangential stress $\sigma_{tt}$ giving rise to surface waves, (whereas 
$\sigma_{nn}= \sigma_{n t} = \sigma_{t n} = 0$ at the boundary )\cite{paoAndmow,achenbach}.  

We are interested here in solutions of the wave equations
\refeq{pde} in bounded domains in two dimensions. The set of
eigenfrequencies is discrete and the solutions depend on the shape of the
domain and on the boundary conditions. The wave equation is non-separable
for typical domain shapes and numerical schemes, as for example boundary
element methods (BEM), have to be employed to calculate the eigenfrequencies and
corresponding wave-functions. The BEM is typically less straightforward in the
elastodynamical case compared to applying it to the scalar Helmholtz equation.
The integral kernels become hyper-singular for common boundary conditions as for
example free boundaries and the displacement field is vectorial. Standard
techniques to apply BEM to the Navier-Cauchy equations are described in
\cite{bonnet}.

Relatively simple expression for the boundary integral kernels can, however,
be obtained when considering the high frequency limit. A generalization of
Bogomolny's transfer operator method \cite{bogomolny}, derived originally for the
Helmholtz equation in bounded domains, yields in two dimensions a boundary kernel
in the form of a $2\times 2$ matrix, that is,
\be{BogoT}
T(q,q'; \omega) = \sqrt{\frac{\omega}{2 \pi \rm i}} \,
\sqrt{\left|\frac{\partial^2 L}{\partial q\, \partial q'}\right|}\,
\MatrixII{\alpha_{pp}} {\alpha_{ps}} {\alpha_{sp}} {\alpha_{ss}}
\MatrixII {\sqrt{\frac{1}{c_p}} e^{i k_p L(q,q') - i\nu_p \frac{\pi}{2}}}
{0} {0}
{\sqrt{\frac{1}{c_s}} e^{i k_s L(q,q') - i \nu_s \frac{\pi}{2}}}
\ee
where $q$, $q'$ denote points on the boundary of the domain.
 $L(q,q')$ is the distance between $q$ and $ q' $ in the
two-dimensional $(x,y)$ - plane and the reflection coefficients are
defined in \refeq{refl-coef}. The additional phases $\nu_{p,s}$
are Maslov indices which count the number of caustics along the path for
each polarization \cite{gutbook}.  Approximations to the eigenfrequencies
are then obtained by solving
\be{quantT}
\det\left({\bf 1} - {\bf T}(\omega)\right) \stackrel{!}{=} 0 \, .
\ee
The transfer operator \refeq{BogoT} can be viewed as a discrete wave
propagator acting on boundary wave functions by mapping outgoing
two-component wave vectors at a point
$q$ on the boundary into outgoing wave vectors at $q'$. Wave mixing
at the boundary enters through the matrix ${\bm \alpha}(q)$.
Snell's law \refeq{Snell} is obtained by considering the
two-step operator
\[
T^2(q,q''; \omega) = \oint dq' T(q,q'; \omega) T(q',q''; \omega)
\]
in stationary phase approximation. Considering $n$ step operators
$T^n$, one can derive periodic orbit trace formulas as presented in
\cite{couch}, see also section \ref{sec:sec3}.

The transfer operator is in many respect
a fairly crude approximation of the true BEM. It is the leading order
term in a $1/\omega$ expansions of the exact boundary integral kernel and
does in particular not contain evanescent contributions. It can therefore
not reproduce boundary effects like surface waves as well as
diffraction or higher order mode mixing corrections. It is, however,
a natural starting point to investigate the connection between
the wave dynamics in elastic media of finite size and an underlying
ray dynamics which includes ray splitting.

In the following, we shall study the billiard with probably the most simple
geometry, namely an elastic disc. Even though the wave equation is
separable for this particular shape, there is some degree of wave chaos in
this system, which can be traced back to chaotic components of an
underlying ray dynamics.

\section{Classical ray dynamics for circular domains}
\label{sec:sec2a}
In this section, we shall discuss in more detail
the ray dynamics in isotropic media of general shape
and in particular for two-dimensional circular domains.

We will adopt the following convention for a
ray-trajectory in an elastic isotropic medium:
a trajectory is at any instant in time given by
its position and momentum as well as its polarization
being either of $p$ or $s$ type. We may identify the
wave numbers $k_{\pi} = \omega/c_{\pi}$, $\pi \in \{p,s\}$
as the momenta of polarization $\pi$ with $k_s = \kappa k_p$.
A ray travels along straight lines between impacts with the boundary.
At the boundary, it stays in a given polarization or undergoes mode
conversion with probability
\be{prop}
t_{pp} = t_{ss} = |\alpha_{pp}|^2,\; t_{ps} = t_{sp} = |\alpha_{ps}|^2,
\ee
where the reflection coefficients are given in (\ref{refl-coef}).
The angle of reflection depends on whether mode conversion takes
place or not via Snell's law and so does the momentum.
A trajectory
is uniquely determined after fixing its initial position and
momentum and an infinite sequence of polarizations
$\pi_1, \pi_2, \ldots$, $\pi_i \ldots \in \{s,p\}$
reflecting the probabilistic nature of the dynamics.
The dynamics in an elastic isotropic medium is thus
taking place on two different energy sheets with energies
$E_p = k_p^2$ and $ E_s = k_s^2$ with $E_p < E_s$.
The dynamics on each sheet is deterministic, jumps from
one sheet to the other may occur at the boundaries.

We shall now turn to the dynamics in a circular disc with radius $a$.
The angular momentum $L$ is conserved at impact with a boundary
both for rays staying in a given polarization and those undergoing mode
conversion which follows directly from \refeq{Snell}, that is,
\be{angular-mom}
L = z_p \sin \theta_p = z_s \sin \theta_s = const \, ,
\ee
where we set $ z_p = a k_p$ and $z_s = a k_s$.
The maximal angular momentum possible for fixed frequency
$\omega$ is $|L_{\rm max}| = z_s$, no
wave splitting occurs for $z_p \le |L| \le z_s$.
The dynamics in a disc follows a simple scaling relation and
can be characterized in terms of the dimensionless impact parameters
$b_{\pi} = L/z_{\pi}$ with $b_p = \kappa b_s$.
The mode splitting regime is characterized by $|b_p| <1$, pure $s$ - rays
exist for $1 < |b_p| < \kappa$.
A trajectory takes on at most two different angles of reflection
$\theta_p$ and $\theta_s$ with $\sin\theta_{\pi} = b_{\pi}$.

In the language of dynamical systems theory, one may
say that the dynamics on each energy sheet is integrable, that is,
trajectories for fixed $L$ and a given polarization are confined to
a two-dimensional manifold in phase space with the topology of a
torus. Mode conversion couples two specific tori characterized by
$(E_p, L)$ and $(E_s, L)$ and transitions are possible between
these two tori only, see \reffig{energfig} a. The total dynamics
is thus not ergodic.

For $|b_p| <1$, the same initial condition in phase space does,
however, lead to an exponentially increasing number of possible solutions
due to the transitions between the energy sheets.
This leads, for example, to an exponential increase in the number
of periodic ray trajectories with increasing length, a phenomenon usually
associated with classical chaos. The latter follows directly from the
periodic orbit condition
\be{po}
n_p \Delta \varphi_p + n_s \Delta\varphi_s = 2 \pi m,
\quad n_p + n_s = n \ge 2 m\,  .
\ee
Here, $\Delta \varphi_{\pi}$ is the change in the azimuthal angle
for a ray with polarization $\pi$ between two reflections.
The azimuthal angle and the angle of incidence are related through
the relation $\Delta \varphi_{\pi}  = \pi - 2\theta_{\pi}$. The integer
indices $n_{\pi}$ correspond to the number of ray segments with
polarization $\pi$ and $m$ is the number of rotations around the center.
The number $N(n)$ of periodic orbits with $n$ reflections (including
permutations of the polarizations) increases thus exponentially like
\[ N(n) \sim \frac{n}{2} 2^n \, .\]
The fact, that these orbits (apart from the $m=0$ case) all form
continuous families is, however, a feature known from integrable dynamics.

The dynamics becomes particular simple when we
restrict our considerations to the motion in the radial coordinate only.
The dynamics in $r$ is one-dimensional in each sheet and bound away from the
center due to the centrifugal potential $L^2/2 r^2$, see \reffig{energfig} a.
Transitions between the sheets occur at the boundary $r = a$ for
$|b_p|<1$. The boundary map for the
radial dynamics for fixed $b_p$ is thus a simple stochastic process which
may be described in terms of a graph with two loops of the form shown in
\reffig{energfig} b. The transition rates (\ref{prop}) which again depend
only on the parameters $b_p$ and $\kappa$ define a Markov process
on this graph with topological entropy $h_t = \log 2$ and exponential
decay of correlation for $t_{pp} \ne 0$ or 1. The chaotic component
of the dynamics in the elastic disc is thus a two-level stochastic process.
\FIG{
\rotatebox{0}{\includegraphics[height=5.5cm]{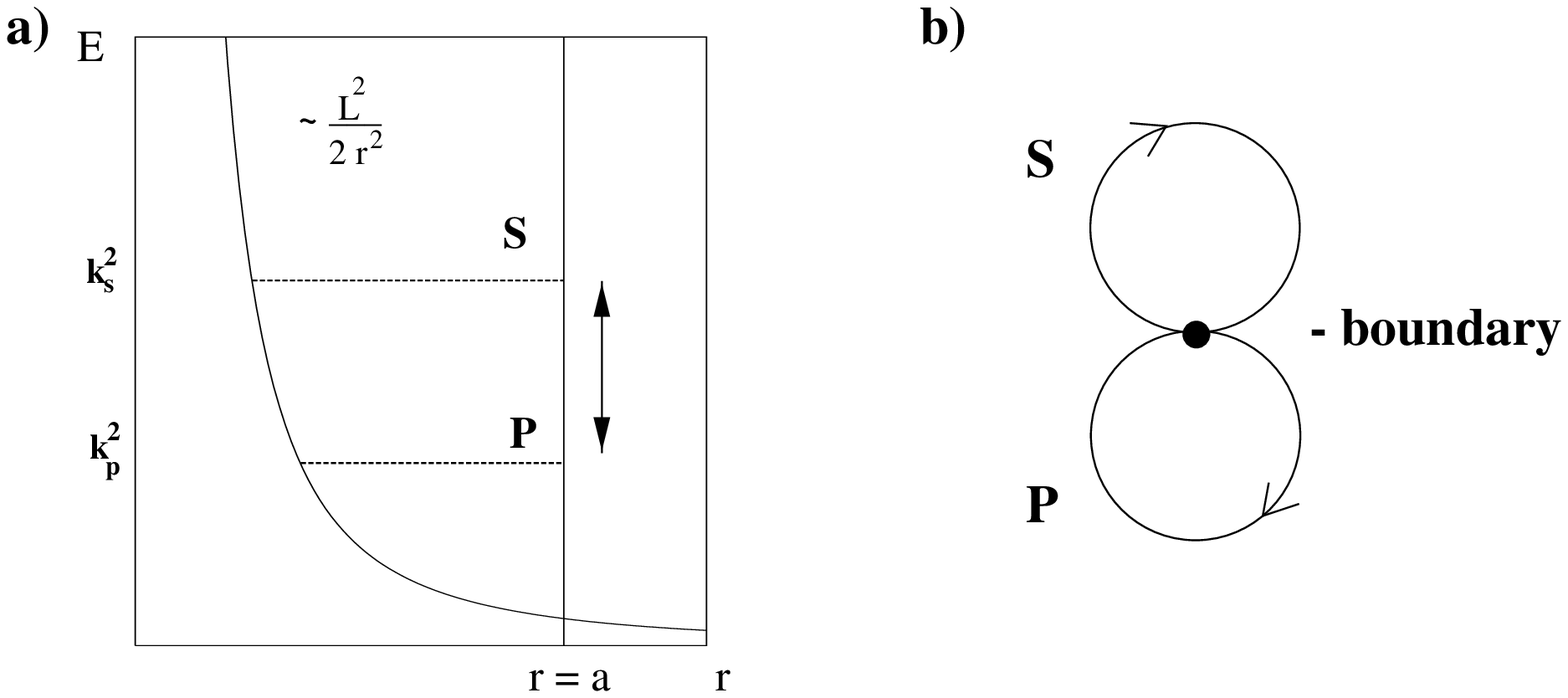}}
}{energfig} { a) Radial dynamics in the disc takes place on two energy sheets
$E_p = k_p^2$ and $E_s = k_s^2 = \kappa^2 E_p$; transitions take place at the
boundary $r = a$.\\
b) The boundary map is equivalent to the probabilistic
dynamics of a Markov process on a binary graph.}
{energfig}

\section{The elastic disc - exact results and the high-frequency limit}
\label{sec:sec2}
The wave equation for a disc of radius $a$ is separable
independent of the boundary conditions. We will briefly discuss the exact
solutions in the case of free boundaries and make a connection between the
eigenfrequencies of the interior problem and the spectrum of the scattering
matrix for the corresponding exterior problem. Details are referred to the
appendices \ref{s-resoExact} and \ref{s-Smatrix}. It will in particular be
shown that the scattering matrix is equivalent to the transfer operator
\refeq{BogoT} in the high-frequency limit.

\subsection{The scattering matrix}
The elastic wave equation can for circular domains in two dimensions be solved
in terms of the basis functions
\be{basis} \psi_l(r,\varphi) = J_l(k_{\pi}r) e^{i l \varphi}\, ,\ee
where $J_l(k_{\pi}r)$ is the $l$th order Bessel function 
with $\pi = p$ or $s$ and $\varphi$ is the azimuthal angle.
The separability of the
wave equation reflects the conservation of angular momentum in the classical ray
dynamics; we obtain as usual that the angular momentum takes on only integer
values $l$.

Applying free boundary conditions to a displacement wave vector obtained from
the potentials \refeq{basis} for fixed $l$ leads to the secular equation,
see appendix \ref{s-resoExact},
\be{secularI}\det({\bf t}_l)(\omega) = 0 \ee
where the traction matrix $\bf t$ is given as
\be{regMatI}
{\bf t}_l
 = \MatrixII{(l^2  - \frac{1}{2} z_s^2) J_l(z_p  ) - z_p J_{l}^{'}(z_p)}
{i l ( J_l(z_p  ) - z_p    J_{l}^{ '}(z_p  ))}
{i l (J_l(z_s  ) - z_s   J_{l}^{ '}(z_s  ))  }
{z_s   J_{l}^{ '}(z_s  )-(l^2  - \frac{1}{2}  z_s^2) J_l(z_s  ) }
\ee
and $z_{\pi} = k_{\pi} a$. The condition
\refeq{secularI} can be rewritten in terms of the scattering matrix for the
outside problem, that is, for in-plane wave - scattering in an infinite plate
with a circular hole. This connection is known as the inside-outside duality
\cite{smilanskyCourse} between eigensolutions of the interior problem and transparent
scattering solutions.  The scattering matrix for fixed angular momentum $l$
is given as \cite{paoAndmow}
\be{scat-def}
{\bf S}_l = - {\bf t}_l^{-} \cdot ({\bf t}_l^{+})^{-1} \, ,
\eeq
see also appendix \ref{s-Smatrix},
where ${\bf t}_l^+$, ${\bf t}_l^-$ are obtained from the traction matrix
\refeq{regMatI} by replacing the Bessel function and its derivatives by
incoming and outgoing Hankel functions. Using
$J_l(z) = (H^{(+)}_l(z) + H^{(-)}_l(z))/2$ which implies the same
identity for the corresponding traction matrices, that is,
\beq
{\bf t}_l= \frac{1}{2}({\bf t}_l^{+} + {\bf t}_l^{-}) \, ,
\eeq
the eigenfrequency condition \refeq{secularI} can be written as
\bea
0 = \det({\bf t}_l) &=& \frac{1}{4} \det({\bf t}_l^{+}) \,
\det({\bf 1} +  {\bf t}_l^{-} \cdot ({\bf t}_l^{+})^{-1})  \continue
    &\equiv& \frac{1}{4} \det({\bf t}_l^{+}) \, \det({\bf 1} - {\bf S}_l).
\label{smat}
\eea
The zeros of the first factor in \refeq{smat} are related to the
resonances for exterior scattering at a
circular cavity and are all in the lower complex $z$ - plane.
An eigenfrequency
for the interior problem of the disc implies that the scattering matrix at
the same frequency has an eigenvalue 1. That is, a scattering solution
exists for which the obstacle, here the disc, is transparent. This principle
holds for general shapes \cite{smilanskyCourse}.

\subsection{The scattering matrix in the high frequency limit}
\label{sec:sec2.2}
In the following we shall derive an approximation to the scattering
matrix in the high frequency limit. 

\paragraph{The mode conversion regime $|b_p| < 1$:}
In the energy - angular momentum regime for which the impact parameter
$|b_p| = |L|/z_p < 1$, the Hankel functions entering ${\bf t}^{\pm}$ may be
split in terms of phases and amplitudes using the
oscillatory Debye approximation;
one obtains to leading order (\refappe{s-Smatrix}),
\be{stat-phas}
{\bf S}_l \approx \MatrixII{e^{-i \phi_p}}{0}{0}{e^{- i \phi_s}}
\cdot {\bm \alpha}_l \cdot \MatrixII{e^{-i \phi_p}}{0}{0}{e^{- i \phi_s}}
\eeq
with
\be{phase}
\phi_{\pi} = z_{\pi} \left( \sqrt{1-b^2_{\pi}} -
b_{\pi} \arccos(b_{\pi})\right) - \frac{\pi}{4}
\ee
and $\bm{\alpha}_l$ is the unitary matrix of reflection coefficients
defined in \refeq{refl-coef} with angles of incidence fixed by the
angular momentum condition \refeq{angular-mom} with $L = l$.
We thus obtain for transitions between polarizations $\pi$, $\pi'$
\be{s-l}
{\bf S}_l(\pi \rightarrow \pi')
\approx \alpha_{\pi \pi'} \cdot e^{- i (\phi_\pi + \phi_{\pi'})} \, .
\eeq
Note that the unitarity of the $S$ - matrix is preserved in this
approximation. This is not true in general for short wavelength
approximations and results here from the quasi one-dimensionality of
the dynamics.

\paragraph{Regime of no mode conversion $ 1< |b_p| < \kappa$ :}
For angular momenta corresponding to incident angles larger than the
critical angle or $|b_p|  > 1$, a somewhat different treatment needs to be
employed.  Here the exponential Debye expansion must be
used for the pressure wave leading to an $S$ - matrix of the form
\be{s-l-s}
{\bf S}_l \approx \MatrixII{1}{0}{0}{\alpha_{ss} e^{-2 i \phi_s}}
\eeq
with a reflection coefficient (in agreement with the plane interface result)
\beq
\alpha_{ss} = - \frac{  Z^*_l}{ Z_l}
\label{alphaBeyond}
\eeq
and
\beq
Z_l = 1 + \cos 4\theta_s + i \,  8 \cos \theta_s \sin^2 \theta_s
\sqrt{\sin^2 \theta_s - 1/\kappa^2} \, .
\eeq
Here, the boundary conditions lead to a pure phase shift dependent on the
angle of incidence. There is no attenuation associated with this reflection
contrary to the wave splitting case. The phase shift is due to a coupling to
a surface longitudinal wave \cite{bedford}.

\paragraph{The transfer matrix for $|b_p| < 1$:}
By parametrising the boundary in terms of the azimuthal angle $\varphi$, one
obtains for the transfer operator \refeq{BogoT} in a circular domain
\be{BogoTdisc}
T(\varphi,\varphi'; \omega) = \sqrt{\frac{\omega}{2 \pi i}} \,
\sqrt{\frac{a}{2} \sin\frac{\Delta\varphi}{2}}\,
\bm{\alpha}(\Delta\varphi)
\MatrixII {\sqrt{\frac{1}{c_p}} e^{i k_p d}} {0} {0}
{\sqrt{\frac{1}{c_s}} e^{i k_s d}},
\ee
with $\Delta\varphi =|\varphi-\varphi'|$ and
$d$ is the distance between two points $\varphi$, $\varphi'$ on the boundary,
that is,
\[ d(\varphi,\varphi') = 2 a \sin\frac{\Delta\varphi}{2}\, . \]
The transfer operator depends only on the difference
$\Delta\varphi$ and block-diagonalises with respect to the
Fourier basis $|l> = \exp(i l \varphi)/\sqrt{2 \pi}$ for integer $l$.
One obtains after evaluating the second integral by stationary phase
\be{T-l}
<l|{\bf T}|l'> = \delta_{ll'}\, \bm{\alpha}_l \,
\MatrixII{e^{2 i \phi_p}} {0} {0}
{e^{2 i \phi_s}} \, ,
\eeq
where the phases
\be{ph-trans} 
\phi_{\pi} = \frac{1}{2} k_{\pi} d(\Delta\varphi_{\pi}) -
l\frac{\Delta\varphi_{\pi}}{2}  - \frac{1}{4}\pi
\eeq
are taken at the stationary phase point
\be{ang-cond}
a k_{\pi} \cos \frac{\Delta\varphi_{\pi}}{2} = a k_{\pi} \sin \theta_{\pi} = l
\ee
which is the angular momentum condition \refeq{angular-mom}. 
The phases $\phi_{\pi}$ in \refeq{ph-trans} coincide with \refeq{phase} after
inserting the stationary phase condition 
$\Delta\varphi_{\pi}/ 2 = \arccos b_{\pi}$, Eq.\ \refeq{ang-cond}.
The S-matrix in the approximation \refeq{stat-phas} is thus equivalent
to the hermitian conjugate of the T-matrix up to a simple transformation
in terms of a unitary diagonal matrix. In particular the eigenfrequency 
conditions $\det ({\bf 1}-{\bf T}) \approx \det ({\bf 1}-{\bf S})=0$ 
coincide in the high frequency limit.

\paragraph{The mean density of eigenfrequencies for fixed $l$:}
The mean density of eigenfrequencies $\overline{d}_l$ for fixed $l$ can be
obtained from the scattering matrix ${\bf S}_l$ \cite{smilanskyCourse},
that is,
\[
\overline{d}_l (k_p) = \frac{1}{2 \pi i} 
\frac{d}{dk_p} \log \det({\bf S^{\dagger}}_l) \, .
\]
Inserting the high frequency limit of the $S$ matrix \refeq{stat-phas}
for $|b_p| < 1$ one obtains to leading order
\be{mean-dens}
\overline{d}_l (k_p) =
 \frac{1}{\pi} \frac{d}{dk_p} \left(\phi_p + \phi_s \right)
= \frac{a}{\pi} \left( \sqrt{1 - b_p^2} + \kappa \sqrt{1 - b_s^2} \right) \,
.
\ee
Note that the mean level density depends on $k_{\pi}$ and $l$ only via the
impact parameters $b_{\pi}$. Equivalently, we obtain from \refeq{s-l-s}
for $1 < |b_p| <\kappa$
\be{mean-dens1}
\overline{d}_l (k_p) =
 \frac{1}{\pi} \frac{d}{dk_p} \phi_s 
= \frac{a}{\pi} \kappa \sqrt{1 - b_s^2} \, .
\ee
 
\FIG{
\rotatebox{-90}{\includegraphics[height=9cm]{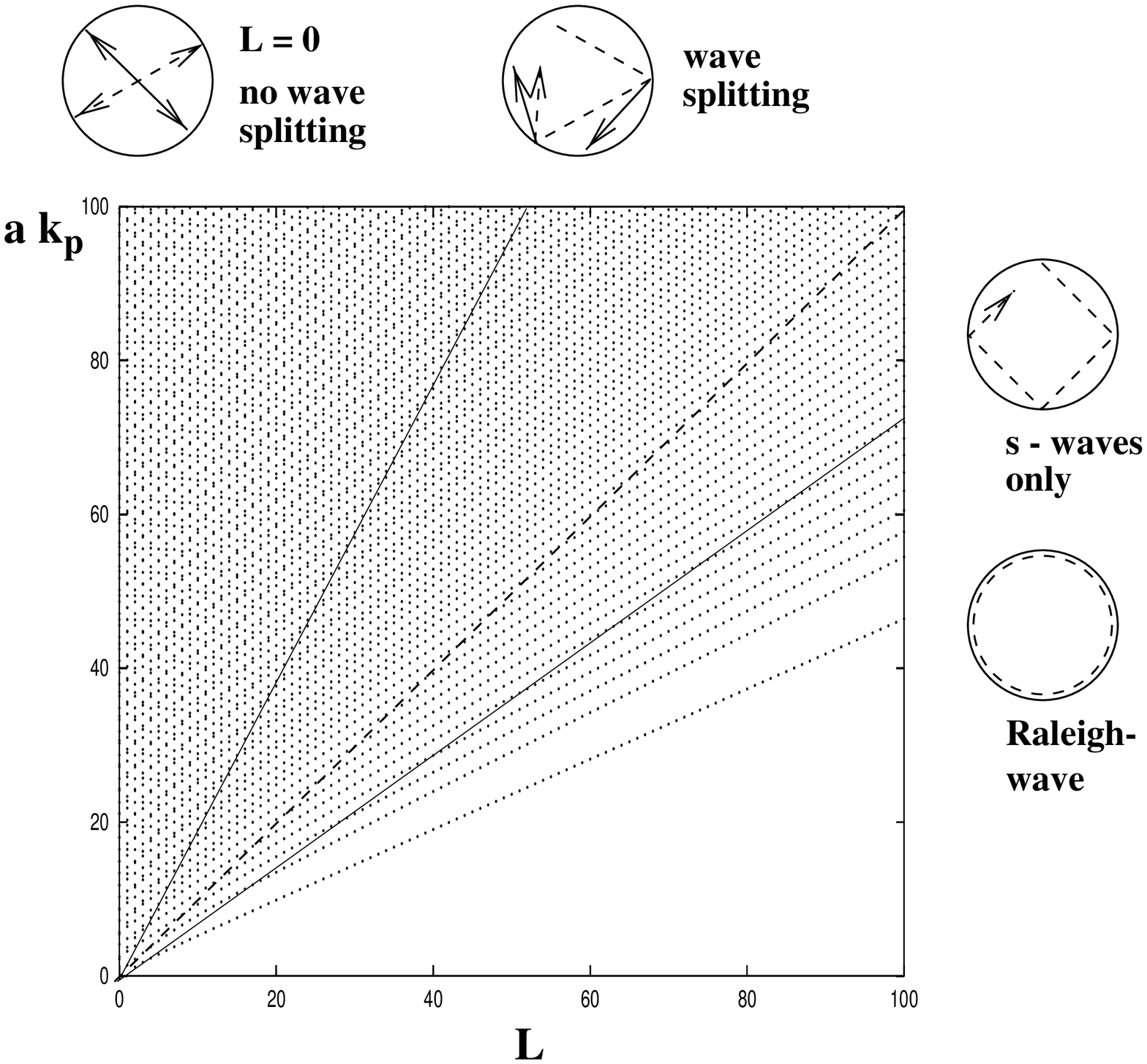}}
}{spec}{The eigenspectrum of the disc for aluminum with $\kappa = 2.05$;
the wavenumber $z_p = a k_p$ of an eigenfrequency is plotted versus the 
angular momentum $l$. Eigenfrequencies with $a k_p < l$ (or $|b_p| > 1$) 
are pure shear states, mode mixing occurs for $a k_p>l$. The lowest states 
in each $l$ - series are due to surface waves (Rayleigh - waves).
}
{spec}

\subsection{Statistical properties of the eigenspectrum}
We saw in the last section that a transition takes place at
$|b_p| = |L|/k_p a = 1$ between a pure shear
wave regime with $|b_p| > 1$ and a mode mixing regime with $|b_p| <1$. This
transition is reflected in the ray dynamics, see section \ref{sec:sec2a}.
Whereas only one family
of (shear-) trajectories exists for fixed $b_p$ with $|b_p| > 1$, there are
infinitely many such families for $|b_p|<1$ and their number
increases exponentially with the length of the trajectories. Such a phenomenon
is reminiscent to the behavior typically found for chaotic classical dynamics.
It was observed in section \ref{sec:sec2a} that the dynamics in the
mode mixing regime can indeed be described by a stochastic Markov process on a
two-loop graph, see \reffig{energfig} b. In the limit $b_p \to 0$, however, the
transition rates $t_{pp} = t_{ss}$ approach one and the two modes decouple
again leaving only two possible ray-trajectories.

All these regimes
should manifests itself also in the spectrum of the elastic disc. Spectral
correlations are known to be particularly sensitive to the degree of chaos
present in an underlying classical (ray-) dynamics \cite{Meh91, Guhr98}. A
popular statistical measure is the so-called nearest neighbour spacing
distribution $P(s)$ giving the probability of finding two adjacent
eigenvalues of the spectrum (unfolded to mean level separation one) a distance
$s$ apart. $P(s)$ follows a Poisson distribution for completely
uncorrelated spectra, but has been conjectured to coincide with the results
obtained for ensembles of random Hermitian matrices for completely chaotic
dynamics.

In \reffig{spec}, the spectrum obtained from the exact eigenfrequency
condition \refeq{regMatI} for an aluminum disc with $\kappa = 2.05$ is
shown. Here, the wavenumber $z_p = a k_p$ of an eigenvalue with angular
momentum $l$ is plotted. One notices a difference in the mean density of
eigenvalues  for fixed $l$ above and below the diagonal $|b_p| = |l|/z_p = 1$,
see Eqs.\ \refeq{mean-dens} and \refeq{mean-dens1}. The lowest eigenfrequency
in each $l$ - series can be attributed to a surface (or Rayleigh-) wave.
Mode mixing occurs for $|b_p| < 1$ or $a k_p > L$. 

First we look at $P(s)$ for the total spectrum which is obtained by projecting
the different $l$ - eigenfrequency series in \reffig{spec} onto the $z_p$-axis.
One finds indeed good agreement with Poisson statistics as shown in
\reffig{nns} a. This reflects the fact that the ray-dynamics in the elastic
disc is not ergodic but restricted to manifolds with fixed angular momentum
$L$ which have the form of a single torus for $|b_p|>1$ or two coupled tori for
$|b_p| < 1$. The eigenfrequencies for different $l$ - series are thus
uncorrelated, which leads to vanishing correlations in the full spectrum for
large $z_p$ after projection onto the $z_p$-axis. \\

In order to see the influence of wave mixing on the spectrum, we need to
study spectral correlations within a given $l$ series. Indeed, the
wave - dynamics for fixed $l$ is given by the $2 \times 2$ transfer matrix
Eq.\ (\ref{T-l}) (or the equivalent matrix for $|b_p| >1$ obtained from
\refeq{s-l-s}) being of the form
\beq
{\bf T}_l(z_p) =
{\bm \alpha}_l(b_p) \cdot
\MatrixII{e^{i z_p\tilde{\phi_p}(b_p)}}{0}{0}
{e^{ i z_p \tilde{\phi_s}(b_p)}} \, .
\eeq
The transfer matrix ${\bf T}_l$ consisting of a unitary transition
matrix $\bm \alpha$ times a diagonal matrix is typical for propagation
on quantum graphs as studied in \cite{Kot97,Tan00}. The phases in the diagonal
matrix $\tilde{\phi_p}, \tilde{\phi_p}$ are thereby interpreted as the
lengths of the bonds in the graph, here the two loops in \reffig{energfig} b.
As we change $z_p$, we expect the correlations within a given $l$ series
to change according to the degree of wave-mixing possible, that is, to
the degree, that $t_{pp}$ deviates from zero or one. 
Note that the transition amplitude $t_{ps}$ allowing for transitions between $p$ and
$s$ waves goes to zero both in the limit $b_p \to 0$ and $b_p\to 1$, see
Eq.\ \refeq{refl-coef}; the wave modes decouple in these limits.
We expect
furthermore, that correlations are locally the same for different $l$ series
with $b_p$ fixed, that is, along straight lines in \reffig{nns} with
$b_p = const$. We therefore study the nearest neighbour spacing
distribution $P(s)$ for eigenfrequencies lying in a window given by the
intersection of the cone $b_p \pm \Delta b_p$ and the line $L=l=const$; the
observed distributions $P(s, b_p)$ are indeed independent of $l$ and
we may average $P(s,b_p)$ over different $l$ - series.

The result is shown in \reffig{nns} b for three different values of $b_p$
and $\Delta b_p = 0.01$. In addition, the value of $t_{pp}$ as a function
of $b_p$ is given; maximal mixing corresponds to $t_{pp} = 0.5$. The
distributions are neither similar to the Poisson distribution, nor to any
of the random matrix distributions. This non-universal behavior is typical
for graphs with only a few bonds \cite{Bar00}. One notices, however, that
a gap is opening up in $P(s)$ for small $s$ as $b_p$ increases from zero.
This can be interpreted as level-repulsion due to mode-mixing which
increases as $t_{pp}$ deviates from one. For $b_p \to 1$ a different
effect sets in; the pressure mode is suppressed and we witness the
transition form a two mode to a one mode wave dynamics with equidistant
eigenfrequencies.

\FIG{
\rotatebox{-0}{\includegraphics[height=13cm]{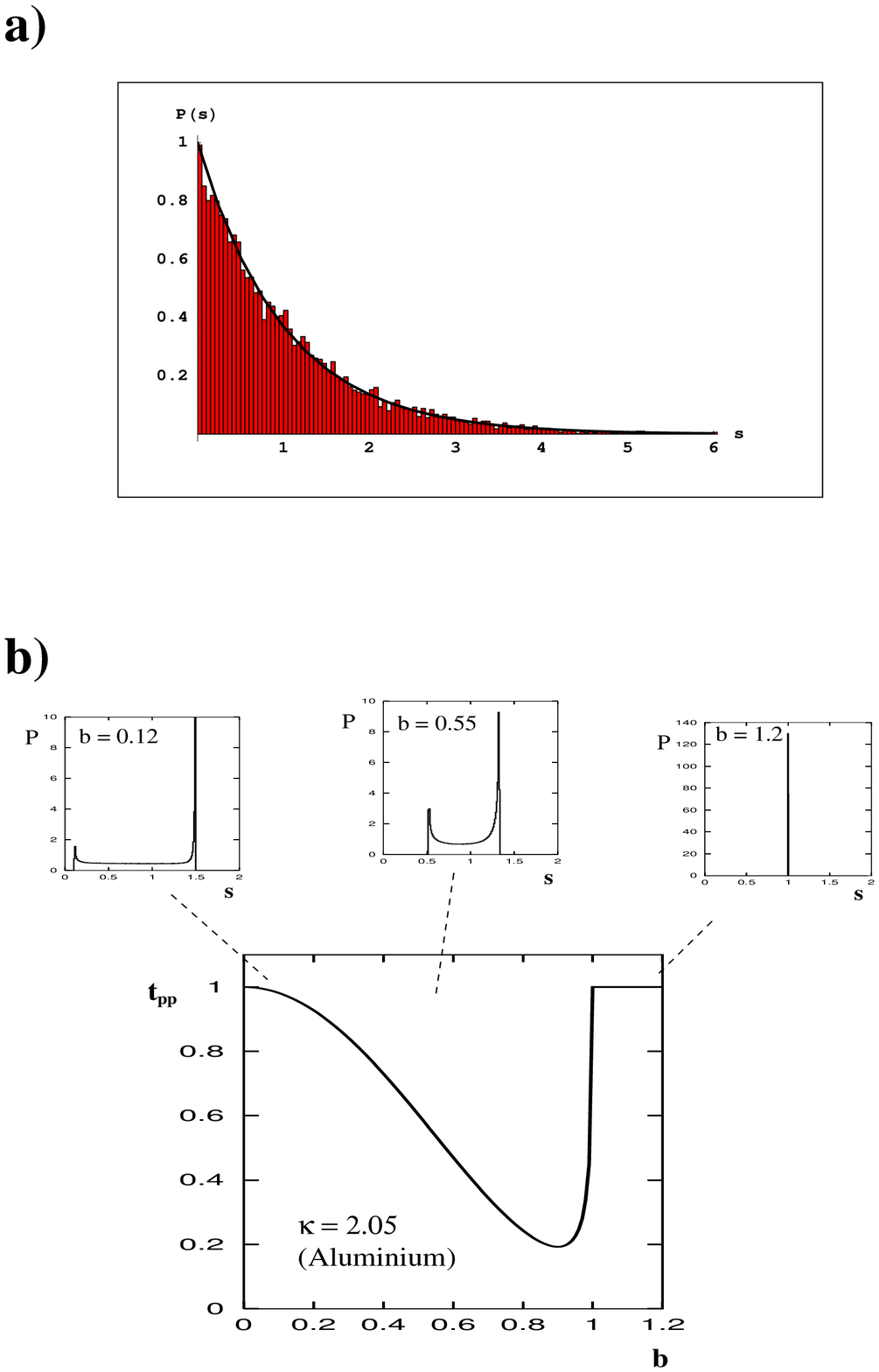}}
}{nnsb}{Statistical properties of the eigenspectrum for aluminum with
$\kappa = 2.05$. a) the nearest neighbour spacing distribution (NNS) for
all levels; b) NNS distributions for fixed impact parameter $b_p$
together with the transition probabilities $t_{pp}$ as a function
of $b_p$. The NNS is obtained for stretches of eigenvalues lying in a range
$b_p \pm \Delta b_p$ for angular momenta from $l = 300 - 3000$;
}
{nns}

\section{Eigenfrequency density and a periodic orbit trace formula}
\label{sec:sec3}
So far we have shown that the traction matrix as well as the
scattering and transfer matrix can be brought into block-diagonal
form where each  $2 \times 2$ block produces the spectrum for fixed
angular momentum $l$. In this section, we will make an explicit
connection between the full spectrum and periodic trajectories
in the elastic disc by looking at the total spectral density
\[ d(\omega) = \sum_i \delta(\omega - \omega_i) \, ,\]
where the sum runs over all eigenfrequencies (in terms of the 
angular velocity $\omega$) of the disc. 

The spectral density can quite generally be written in terms
of a smooth part and oscillatory contributions, the latter
containing the period orbit contributions, that is,
\[ d(\omega) = d_{smooth}(\omega) + d_{osc}(\omega) \, . \]
 
The smooth part gives the mean density of states which may be obtained using
\[
d_{smooth}(\omega ) = \frac{d N_{smooth}}{d \omega }
\]
where $N_{smooth}$ is the mean part of the spectral counting function 
$N(\omega)$ giving the number of levels below $\omega $. General 
results for the smooth part of the counting
function of isotropic elastic media with free boundary conditions
can be given \cite{SAFAROV,Yale}, that is,
\beq
N_{smooth}(\omega) \approx \frac{c_p^{-2} + c_s^{-2}}{4 \pi} \, 
S \, \omega^2 + \frac{\beta}{4 \pi c_s} \,  L \, \omega + \mathrm{o}(\omega)
\label{Nsmooth}
\eeq
with
\[
\beta =  4 \eta - 3 + \kappa + 4 I/\pi
,\] and
\[
I = \int_{1/\kappa}^1 \arctan
\frac{(2-1/\xi^2 )^2}{4 \sqrt{\left(1-\frac{1}
{\kappa^2 \xi^2}\right)\left(\frac{1}{\xi^2} -1\right)}} \, d\xi \, .
\]
Here $S,L$ are the surface area and the perimeter of the domain and
finally $\eta = c_R/c_s$ with $c_R$ the Rayleigh wave velocity \cite{lAndl}.
The leading term corresponds to the available phase space volume whereas the
next order term contains corrections due to surface states.
In contrast to the scalar Helmholtz equation for which the
expansion of $N_{smooth}$ can be worked out to arbitrary order in principle
\cite{BerHowls} only the first two terms are known in elastodynamics
at present.

The fluctuating part of the density of states which contains all the information
about individual eigenfrequencies of the interior problem can be written in
terms of the scattering matrix of the outside problem \cite{smilanskyCourse,smilanskyIddo}.
One obtains
\beq
d_{osc}(\omega) =  \frac{1}{\pi} \, \Im  \sum_{n=1}^{\infty} \frac{1}{n}  \frac{d}{d\omega}
\Tr \left({\bf S}^{n}(\omega)\right)^{\dagger} \, .
\label{dOsc}
\eeq
In the high frequency limit this term is  related to periodic ray trajectories in 
the disc as will be shown below.

\subsection{Oscillatory part of the density of eigenfrequencies}
We will first derive a periodic orbit expression for $\Tr {\bf S}^n$ in the high 
frequency limit $\omega \gg 1$ using the block-diagonal form of the scattering matrix and the
approximation \refeq{stat-phas}, that is,
\bea
\Tr {\bf S}^n(\omega) &=& \sum_{|l|\le l_{max}} \Tr {\bf S}^n_l \continue
&\approx&
\sum_{|l| \le l_{max}} \sum_{{\bm \pi}_n}
 {\cal A}_{{\bm \pi}_n}  \,  e^{-i 2 (n_p \phi_p + n_s \phi_s)} \, .
\eea
Here, $l_{max} \sim z_s$ denotes the maximal angular momentum
and the second sum runs over all binary symbol strings
${\bm \pi}_n = \pi_1 \pi_2 \ldots \pi_n$ of length $n$ with $\pi_i \in \{p,s\}$.
The amplitude ${\cal A}_{{\bm \pi}_n}$ is obtained as product over reflection
coefficients \refeq{refl-coef}, that is
\[ {\cal A}_{{\bm \pi}_n}=  \prod_{i=1}^n \alpha_{\pi_i  \pi_{i+1}}\, ,\]
$\phi_{\pi}$ are the phases defined in  \refeq{phase} and $n_p$, $n_s$ equal the
number of times the symbol $p$, $s$ appears in ${\bm \pi}_n$.
The sum over binary symbol strings for fixed $l$ is equivalent to a 
sum over all periodic paths in the binary graph \reffig{energfig}.
Note that the reflection coefficients as well as the phases $\Phi_{\pi}$
depend explicitly on $l$ and $\omega$.
 
Next, we use Poisson summation to write the sum over $l$ as 
\bea
\Tr {\bf S}^n &\approx& \sum_{{\bm \pi}_n} \sum_{|l| \leq l_{max}}
{\cal A}_{{\bm \pi}_n} \, e^{- 2 i (n_p \phi_p + n_s \phi_s)} \continue
 &=& \sum_{{\bm \pi}_n} \sum_{m=-\infty}^{\infty} \int_{-l_{max}}^{l_{max}} dl
{\cal A}_{{\bm \pi}_n} \,
 e^{-2 \, i \, (n_p \phi_p + n_s \phi_s)-2 \pi \, i \, m l} \, .
\label{StatIntegral}
\eea
Evaluating the integrals by stationary phase using
\beq
\frac{d \phi_{\pi}}{dl} = - \arccos(l/z_{\pi}) = - \frac{\Delta \varphi_{\pi}}{2}\, ,
\eeq
where $\Delta \varphi_{\pi} $ is the angle spanned by a ray-segment with
polarization $\pi$ between two reflections, we obtain the stationary
phase condition
\beq
\Delta \varphi_{total} = n_p \Delta \varphi_p + n_s \Delta \varphi_s = 2 \pi m \, .
\eeq
This is precisely the periodic orbit condition \refeq{po}, that is, only those
angular momenta $l^* = z_{\pi} \cos\frac{1}{2}\Delta \varphi$
contribute significantly, for which a periodic orbit exists
at frequency $\omega$.
The second derivative of the phases in \refeq{StatIntegral} is
\beq
 -\frac{d (\Delta \varphi_{total}- 2 \pi m)}{dl}
= 2 \left( \frac{n_p}{z_p \cos \theta_p} +  \frac{n_s}{z_s \cos \theta_s}  \right) \,
\label{fluctuationDet}
\eeq
with $\theta_{\pi}$, the angle of incident and
$\cos\theta_{\pi} = \sin\frac{1}{2}\Delta \varphi_{\pi}$.

After evaluating the phases $\Phi_{\pi}$ in \refeq{StatIntegral} at the
stationary phase point $l^*$, one obtains for the total phase
\beq
2 (n_p \Phi_p + n_s \Phi_s) + 2 \pi l^* m =  n_p k_p d_p + n_s k_s d_s - n \frac{\pi}{2} = 
\omega T - n \frac{\pi}{2}
\eeq
with $n = n_p + n_s$ and $d_{\pi} = 2 a \sin \frac{1}{2}\Delta\varphi$, the
length of a ray-segment of polarization $\pi$ between reflections. Furthermore, 
$T$ is the period of the periodic orbit. We finally obtain
\beq \label{trace-S}
\Tr {\bf S}^n \approx \sqrt{\pi a \omega} \sum_{po}^{(n)} { A_{po}}
e^{-i \omega T_{po} + i n \pi/2 - i \pi/4} \, ,
\eeq
where the sum is taken over all periodic rays with $n$ reflections and
\[
A_{po}= \frac{{\cal A}_{po}}{\sqrt{\frac{n_p c_p} {\cos \theta^{po}_p} +
\frac{n_s c_s }{\cos \theta^{po}_s} }}\, .
\]

By taking the complex conjugate and the derivative with respect to $\omega$ of 
eq.\ \refeq{trace-S}, we finally obtain the spectral density to leading order in 
$1/\omega$ as
\bea
d(\omega) &\approx& \sqrt{\frac{a \omega}{\pi}} \sum_{n=1}^{\infty}
\frac{1}{n}\sum_{po}^{(n)} A_{po} T_{po} \, \cos ( \omega T_{po} - n \pi/2 +  \pi/4) \continue
 &=&  \sqrt{\frac{a \omega}{\pi}}
\sum_{ppo}\frac{T_{ppo}}{\sqrt{\frac{n_p c_p}{\cos \theta^{po}_p} +
\frac{n_s c_s}{ \cos \theta^{po}_s}}}
\sum_{r=1}^{\infty} \frac{{\cal A}^r_{ppo}}{r^{3/2}}
\, \cos( r(\omega T_{ppo} - \hspace{-1mm}n_{ppo} \pi/2)+\pi/4) . 
\label{WaveSplitContrib}
\eea
The last expression is obtained after summing over orbits related by cyclic
permutations of the symbol code ${\bm \pi}$ and the sum is now taken over 
all primitive periodic orbits (ppo) of arbitrary length, that is, over orbits 
not including repetitions and cyclic permutations. The second sum over $r$
then includes the repetitions.\\

We note in passing that the result \refeq{WaveSplitContrib} can also be
derived from  a generalization of the
{\it abelian trace formula} \cite{CreaghLittle,Creagh} valid for systems
with continuous symmetries. Here the symmetry is used to integrate over families
of orbits $\Gamma$. Due to the rotational symmetry in the disc, one obtains
\beq
d_{osc}(\omega) \approx \sqrt{2/\pi}
\sum_{orbit \, \, \Gamma} \frac{{\cal A}_\Gamma T_\Gamma }
{a_\Gamma \sqrt{|\pde \theta/\pde L|}} 
\cos\left(\omega T_\Gamma - \sigma_\Gamma \frac{\pi}{2} - \frac{\pi}{4} \right),
\eeq
where $T_\Gamma$ is the period of the orbit, $\sigma=\mu-1$ with $\mu$ the 
Maslov index, $a_\Gamma$ the order of the (possibly) discrete symmetry group 
of the orbit and finally ${\cal A}_\Gamma$ the product of the plane wave 
reflection coefficients for scattering at the boundary. Finally, 
$\pde \theta/\pde L$, (also called the anholonomy matrix), describing the 
negative change of perimeter angle due to a change of impact parameter is
\[
\frac{\pde \theta}{\pde L} = + 2 \left( \frac{n_p}{z_p \cos \theta_p}
+ \frac{n_s}{z_s \cos \theta_s} \right) ,
\]
in agreement with \refeq{fluctuationDet}.

\subsection{Periodic orbit spectrum}
Eqn.\ \refeq{WaveSplitContrib} gives an explicit connection between periodic ray
trajectories in the disc and the eigenfrequencies of the system. By taking a Fourier
transform of $d(\omega)$ one should be able to recover the periodic
ray solutions including orbits which change polarization along their path. 
To suppress high-frequency oscillations in the signal, we convolute
Eq.\ \refeq{WaveSplitContrib} on both sides by a Gaussian test function as was also
used in \cite{brack}. The smoothing depends on a parameter $\eta$ proportional to the width,
\beq
w(z_p) = \frac{1}{\eta \sqrt{\pi}} \, e^{-(z_p/\eta)^2}, \, \, z_p = k_p a  \, .
\label{testFunction}
\eeq

Figure \ref{f_period} shows a comparison between a numerically calculated period 
spectrum obtained from the first $23000$ eigenvalues of a disc using \refeq{secularI} 
and the approximative result \refeq{WaveSplitContrib}, here for  
polyethylene with $\kappa = 3.61$.
The smooth part of the spectrum is removed and the density of eigenfrequencies is then
Fourier transformed. This obtained period spectrum shows numerous peaks which fall 
roughly into three classes: orbits being of pure pressure, pure shear and mixed 
polarization type. In general the first class consists of the shortest orbits 
since the pressure waves have the fastest velocity. At $t \approx 3.2$ msec, we have an
infinite number of pressure orbits accumulating at the boundary;
orbits of higher winding number like the pentagram at $t \approx 4.9$ msec can also be resolved. 
Next, orbits with segments of both pressure and shear polarization type
arise. Again one finds accumulation towards a limit orbit with the pressure segments 
becoming tangential to the boundary around $t\approx 5.5$ msec. We note here clear 
deviations of the actual numerical period spectrum from periodic orbit theory. 
Shear waves have  incidence angles close to the critical angle and 
surface contributions become relevant.

We note, that for the first time, 
periodic orbits changing polarization along their paths could clearly be identified 
in a Fourier spectrum of an eigenspectrum of an elastic body.
Quite characteristic is the decay of peak height as the orbits increase in length. 
Note, however, that the class of pure shear orbits with long periods show up as 
comparatively high peaks in the spectrum. This is due to the lack of mode-conversion 
when the shear segments turn towards tangential incidence with a reflection coefficient 
becoming a pure phase \refeq{alphaBeyond}.  
Surface orbits like the pure Rayleigh orbit at $ t \approx 12.2$ msec can also 
be identified.

\FIG{
\rotatebox{0}{\includegraphics[height=10cm]{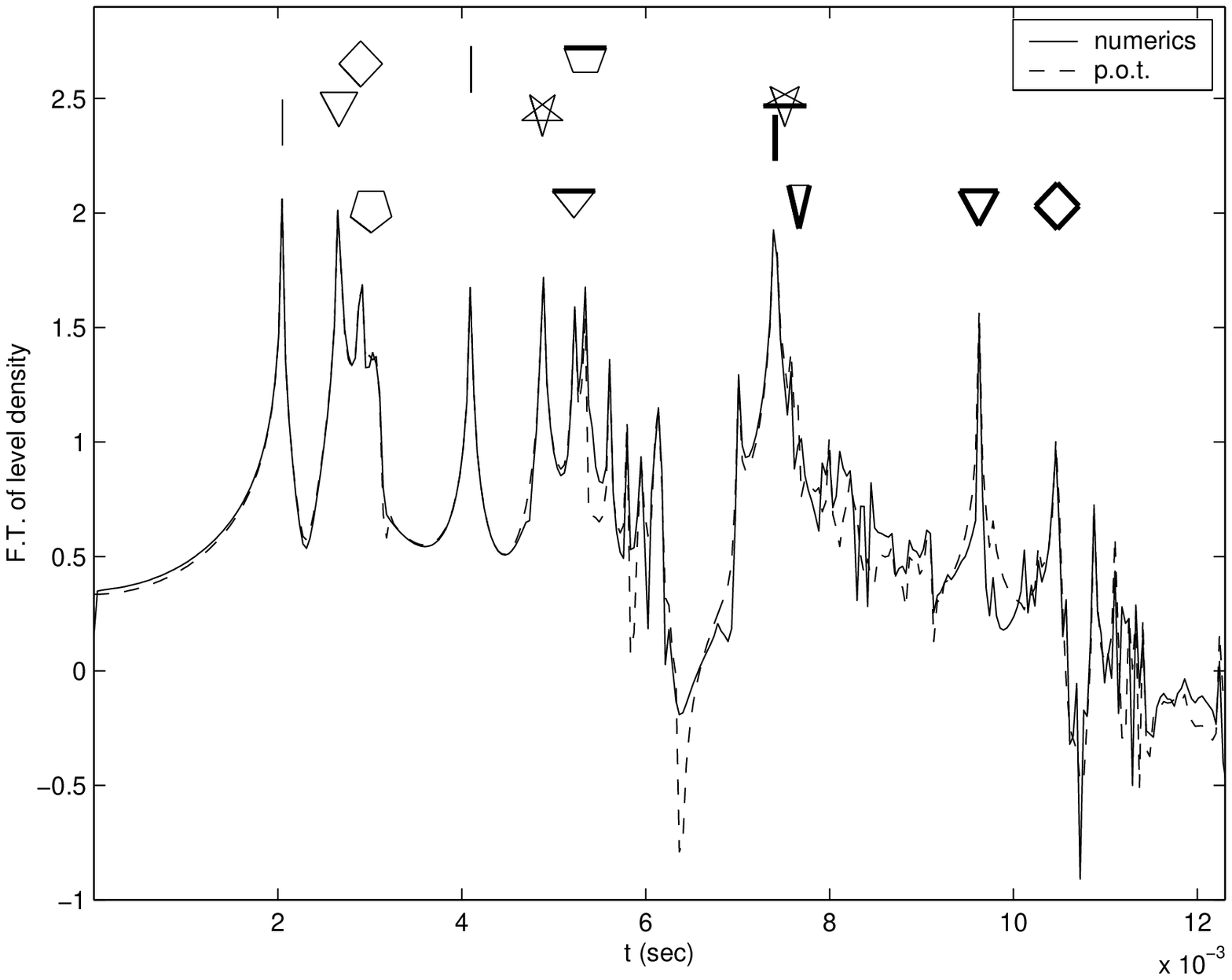}}
}{Period Spectrum}{FFT of oscillating level density. The orbits are depicted with thin/fat
 lines for pressure/shear-polarization and are positioned according to their period. The 
periodic orbit theory (``p.o.t.'') refers to  \refeq{WaveSplitContrib}.  The smoothing 
parameter in \refeq{testFunction} is chosen here as $\eta = 0.2$. The actual material 
corresponds to polyethylene with $c_p = 1950 \, m/s$ and $c_s = 540 \, m/s$. 
Finally the disc radius is $a = 1 \, m$.}
{f_period}

\section{Summary and Discussion}

We have studied the in-plane eigenfrequency spectrum for the
elastic wave equation in two spatial dimensions with circular
boundaries. It was shown that the eigenmodes can be expressed
in terms of periodic rays of an underlying billiard-like classical
dynamics. The ray dynamics conserves angular momentum,
the wave equation becomes separable in polar
coordinates, accordingly. It was pointed out, however, that the existence
of two wave modes with different velocities partially destroys
the integrability of the problem; the ray dynamics
for fixed angular momentum takes place on two different energy
manifolds in phase space for the two polarization. The dynamics
on each manifold is one-dimensional and thus integrable,
transitions between the energy sheets at the boundary introduce
a purely probabilistic component. The classical
dynamics corresponding to the elastic wave equation is therefore
not deterministic and thus not integrable in the sense of
Hamiltonian dynamics.

By solving the wave equation explicitly and deriving high
frequency approximations employing both the scattering matrix and the
transfer operator, a connection between the wave dynamics in the disc
for fixed angular momentum and the unitary propagation on a simple quantum
graph could be established. Spectral correlations due to wave mixing
manifest itself in a gap in the nearest neighbor spacing distribution
and thus strong level repulsion.
Finally, the full level density was expressed in terms of periodic orbits which
could be identified explicitly in the Fourier transform of the exact
density of eigenstates.

The main corrections omitted in the high frequency approximations derived here
occur for nearly tangential orbits both for pressure and shear components.
This regime calls for a more refined approximation of the Bessel functions
occurring in the traction matrix \refeq{regMatI} as for example uniform
approximations. Furthermore, periodic orbits accumulate at the boundary and
the stationary phase approximation used to solve the integrals
\refeq{StatIntegral} breaks down. These corrections give rise to surface waves
typical for free boundary conditions.

Furthermore, higher order terms in the reflection coefficients become
important whenever the leading term $\alpha_{ps}$ vanishes, that is, for
normal incident $\theta \to \pi/2$ corresponding to $L \to 0$. Periodic orbits
at $L=0$ having both $p$ and $s$ segments can indeed been identified
in the Fourier spectrum  \reffig{f_period}. A detailed analysis of
these effects will be discussed elsewhere.\\

\noindent
{\bf Acknowledgment}\\
Financial support by the {\em EPSRC} and the EU - research training
network {\em Mechanics and Symmetry in Europe (MASIE)} is gratefully
acknowledged.
 
\begin{appendix}
\section{In-plane eigenfrequency spectrum for an elastic disc}
\label{s-resoExact}
Due to the rotational symmetry of the disc, the wave equation separates in an angular and radial part. 
The eigenfunctions regular at the origin may be expressed in terms of Bessel functions; one writes the 
displacement field \refeq{helm} as
\[
{\bf u}_p^l = \nabla (J_l(k_p r) e^{-i l \varphi} ) \hspace{1cm} 
{\bf u}_s^l = \nabla \times {\bf \hat{z}} (J_l(k_s r) e^{-i l \varphi} )
\]
and $p$, $s$ refers to  pressure and shear polarization as usual, (also called ``primary'' and ``secondary'' 
wave in seismology referring to the time of arrival of these waves). A general interior eigenfunction can
be  expanded in these displacement fields, that is,
\beq
{\bf u}_l = a_1 \, {\bf u}_p^l + a_2 \; {\bf u}_s^l\, .
\eeq
To fulfill the free boundary condition, the traction of ${\bf u}_l$ has to vanish
at the boundary, that is,
\beq
{\bf t}({\bf u}_l)= a_1 \, {\bf t}({\bf u}_p^l) + a_2 \, {\bf t}({\bf u}_s^l) = 0\, .
\eeq
Writing the traction in terms of its radial and angular direction, one obtains
\bea
0 &=& a_1 \, {\bf t}_r({\bf u}_p^l) + a_2 \, {\bf t}_r({\bf u}_s^l) \continue 
0 &=& a_1  \,{\bf t}_{\varphi}({\bf u}_p^l) + a_2 \, {\bf t}_{\varphi}({\bf u}_s^l) \, .
\eea
Expressed as a  matrix equation this becomes
\beq
(a_1 a_2) \cdot {\bf t}_l = {\bf 0} \, ,
\eeq
where we have collected the coordinates of both polarizations in a matrix
\bea
{\bf t_l} &=& [t^{l}_{\pi i}] \continue 
 &=& \MatrixII{(l^2  - \frac{1}{2} z_s^2) J_l(z_p  ) - z_p J_{l}^{'}(z_p)}{i l ( J_l(z_p  ) - z_p    J_{l}^{ '}(z_p  ))}{i l (J_l(z_s  ) - z_s   J_{l}^{ '}(z_s  ))  }{z_s   J_{l}^{ '}(z_s  )-(l^2  - \frac{1}{2}  z_s^2) J_l(z_s  ) }
\label{regMat}
\eea
with $i = r$ or $\varphi$ and $\pi = p$ or $s$ and thus 
$t_{\pi i}= {\bf t}_i({\bf u}_{\pi})$. We set as usual $z_{\pi} = a k_{\pi}$ 
with $a$, the radius of the disc. A  superposition of these two polarizations 
fulfills the boundary condition only when
\beq
\det({\bf t}_{l}) = 0
\label{regDet}
\eeq
which is the eigenfrequency condition \refeq{secularI}. For more details see 
for example \cite{Kitahara} in which an expression equivalent to 
\refeq{regMat} is derived.

\section{S-matrix}
\label{s-Smatrix}
\subsection{S-matrix in terms of traction operators}
Assume a general boundary condition
\beq
{\bf \Omega} ({\bf u}) = {\bf 0}
\label{bcGen}
\eeq
with $\bf u$ an $n$-dimensional vector wave function and $ \bf \Omega$ a 
linear operator. Denote ${\bf u}_i$ the wave field being non-zero in its
$i$-th component only with $i \in \{ 1, ..., n \} $. 
The scattering process of an incoming pure $i$-wave, ${\bf u}_i^-$, may then be
described as
\be{scat-standart}
{\bf u} = {\bf u}_i^- + \sum_{j} S_{ij} {\bf u}_j^+ 
\eeq
where ${\bf u}_j^+$ denotes outgoing pure polarizations. The Eq.\ 
\refeq{scat-standart} must satisfy the boundary condition \refeq{bcGen}
for all $i$. Solving for the scattering matrix, one obtains
\beq
{\bf S} = - {\bf \Omega (u}^-) \cdot ({\bf \Omega (u}^+))^{-1} \, .
\label{smatGen}
\eeq
The operator ${\bf \Omega (u)}$ may be represented in matrix form
\beq
(\Omega (u_i))_j \, ,
\eeq
where the index $i$ represents the component of the vector field $\bf u$ and 
$j$ denotes the vector component of the operator $\bf \Omega$, 
(that is, $\hat{x}$, $\hat{y}$ or $\hat{r}$, $\hat{\varphi}$ in two
dimensions).

In our case $\bf \Omega (u)$ is given by the traction ${\bf t(u)}$ 
\refeq{freebound} in polar coordinates. The analytic form is obtained from 
\refeq{regMat} using $H_l^{(1)},H_l^{(2)}$ instead of $J_l$.

\subsection{The high frequency limit $\omega \gg 1$ for $b_p = l/z_p < 1$}
Starting from the Eq.\ \refeq{scat-def} expressing the scattering matrix ${\bf S}_l$ in 
terms of traction matrices, we will employ the oscillatory Debye approximation for 
the Hankel functions entering ${\bf t}^{\pm}$, that is,
\beq
H_l^{(1)}(z) \sim \sqrt{\frac{2}{\pi Q}} \,  e^{i \phi}; \quad 
H_l^{(2)}(z) \sim \sqrt{\frac{2}{\pi Q}} \, e^{-i \phi}; \quad 
H_l^{(1) '}(z) \sim i \, \frac{Q}{z} \, H_l^{(1)}(z)  \,
\label{debyeAbove}
\eeq 
with
\beq
Q = \sqrt{z^2 -l^2} \quad \mbox{and} \quad \phi =  Q - \pi/4  -  l \arccos\frac{l}{z} \, .
\label{qAbove} 
\eeq
Here, $z$ is either $z_p$ or $z_s$ with $z_{\pi} = a k_{\pi}$ as usual.
The traction matrices ${\bf t^\pm}$ are obtained from \refeq{regMatI} by replacing 
the Bessel functions in terms of outgoing and incoming Hankel functions, respectively. 
Inserting the Debye approximation and  
separating {\bf t} in amplitude and phase, we obtain \cite{wirzba} 
\beq
{\bf t}^{\pm} = \MatrixII{ e^{\pm i \phi_p}}{0}{0}{ e^{\pm i \phi_s}}  
\cdot {\bf G}^{\pm} \cdot {\bf Z}^{\pm}  \, .
\eeq
Here
\beq
{\bf G}^{\pm} = \MatrixII{\sqrt{\frac{2}{\pi Q_p}}}{0}{0}{\sqrt{\frac{2}{\pi Q_s}}}
\eeq
and
\beq
{\bf   Z}^+ = \MatrixII{(l^2 -z_s^2/2) - 
i Q_p}{i l (1 - i Q_p)}{i l (1 - i Q_s)}{i Q_s -(l^2 -z_s^2/2) } \, ,
\eeq
whereas ${\bf Z}^-$ has the same form as   
${\bf Z}^+$ apart from replacing $Q_{\pi}$ by $-Q_{\pi}$.  Hence,
\beq
{\bf S} = \MatrixII{e^{-i \phi_p}}{0}{0}{e^{- i \phi_s}} \cdot {\bm \alpha}  
\cdot \MatrixII{e^{-i \phi_p}}{0}{0}{e^{- i \phi_s}}
\eeq
with the unitary matrix $\bm \alpha$ defined as 
\beq
{\bm \alpha} = - {\bf G}^- \cdot {\bf Z}^-/({\bf G}^+  \cdot {\bf Z}^+) \, .
\eeq
A straightforward but tedious expansion of ${\bm \alpha }$ in terms of the 
wave numbers $k_{\pi}$, expressing $l$ and $Q_{\pi}$ in terms of the 
incidence angle $\theta_{\pi}$, that is, $l = - z_{\pi} \sin \theta_{\pi}$ 
and $Q_{\pi} = z_{\pi} \cos \theta_{\pi}$, one obtains indeed the reflection 
coefficients \refeq{refl-coef} in leading order. The limit taken corresponds 
to letting $k_\pi \rightarrow \infty$ and $| l |  \rightarrow \infty$ but 
keeping their ratio fixed. As discussed above this ratio corresponds to 
fixed impact parameter/incidence angle.  This finally reproduces 
formula \refeq{stat-phas}. All the formulas are given here for negative 
angular momentum; choosing $l = z_{\pi} \sin \theta_{\pi}$ positive changes
the sign of the off-diagonal components in \refeq{regMatI}. This does 
not alter the eigenfrequency condition \refeq{secularI} reflecting the 
degeneracy of the spectrum with respect to the sign change in $l$.  
\end{appendix}

%

\end{document}